\documentclass[letterpaper, 10 pt, conference]{ieeeconf}  

\IEEEoverridecommandlockouts                              

\usepackage{graphics} 
\usepackage{epsfig} 
\usepackage{mathptmx} 
\usepackage{times} 
\usepackage{amsmath} 
\usepackage{amssymb}  
\usepackage{cleveref}
\usepackage{algorithm}
\usepackage{algpseudocode}
\usepackage{graphicx}
\usepackage{subcaption}
\usepackage{xcolor}
\usepackage{csquotes}

\title{\LARGE \bf
Pathfinders in the Sky: Formal Decision-Making Models for Collaborative Air Traffic Control in Convective Weather
}

\author{Jimin Choi$^{1}$, {Kartikeya Anand}$^{2}$, Husni R. Idris$^{3}$, {Huy T. Tran}$^{4}$, {Max Z. Li}$^{5}$
\thanks{*This work was supported by NASA's Transformative Tools and Technologies Project, Award No. 80NSSC23M0221.}
\thanks{$^{1}$Jimin Choi is with the Department of Aerospace Engineering, University of Michigan,
        Ann Arbor, MI 48109, USA, 
        {\tt\small jiminch@umich.edu}}%
\thanks{$^{2}$Kartikeya Anand is with the Department of Electrical and Computer Engineering, University of Michigan,
        Ann Arbor, MI 48109, USA, 
        {\tt\small kartikan@umich.edu}}%
\thanks{$^{3}$Husni R. Idris is with the Aviation Systems Division, NASA Ames Research Center, Moffett Field, CA 94035, USA, {\tt\small husni.r.idris@nasa.gov}}%
\thanks{$^{4}$Huy T. Tran is with the Grainger College of Engineering,
Department of Aerospace Engineering, University of Illinois Urbana-Champaign, Champaign, IL 61820, USA,
        {\tt\small huytran1@illinois.edu}}%
\thanks{$^{5}$Max Z. Li is with the Departments of Aerospace Engineering, Civil and Environmental Engineering, Industrial and Operations Engineering, University of Michigan,
        Ann Arbor, MI 48109, USA, 
        {\tt\small maxzli@umich.edu}}%
}

\begin{document}

\maketitle
\thispagestyle{empty}
\pagestyle{empty}

\begin{abstract}
Air traffic can be significantly disrupted by weather. \emph{Pathfinder operations} involve assigning a designated aircraft to assess whether airspace that was previously impacted by weather can be safely traversed through. Despite relatively routine use in air traffic control, there is little research on the underlying multi-agent decision-making problem. We seek to address this gap herein by formulating decision models to capture the operational dynamics and implications of pathfinders. Specifically, we construct a Markov chain to represent the stochastic transitions between key operational states (e.g., pathfinder selection). We then analyze its steady-state behavior to understand long-term system dynamics. We also propose models to characterize flight-specific acceptance behaviors (based on utility trade-offs) and pathfinder selection strategies (based on sequential offer allocations). We then conduct a \emph{worst-case} scenario analysis that highlights risks from collective rejection and explores how selfless behavior and uncertainty affect system resilience. Empirical analysis of data from the US Federal Aviation Administration demonstrates the real-world significance of pathfinder operations and informs future model calibration. 
\end{abstract}


\section{Introduction}

\subsection{Background and Motivation}

Convective weather negatively impacts aircraft and flight operations. For a variety of reasons---chiefly, safety and passenger comfort---commercial flights seek to avoid thunderstorms and areas of strong convective activity \cite{gultepe2019review}. Air traffic managers will also proactively close portions of the airspace (e.g., individual airspace sectors and air corridors) in response to current and future severe weather in the forecast \cite{mitchell2006airspace}. Depending on the duration of closure, such air traffic control and flow management actions can negatively impact airline on-time performance, which has a range of economic consequences (e.g., in 2019, the cost of flight delays were estimated by the US Federal Aviation Administration to be in excess of \$33 billion USD \cite{FAA2022}). 

Intuitively, the impact of such closures can be magnified in busy, complex airspace, such as the \emph{terminal airspace} around major airports. The New York terminal airspace, or N90 TRACON, is one particularly poignant example: The in-air entrances into, and exits out of, N90 TRACON are referred to as \emph{arrival gates} and \emph{departure gates} by air traffic control, respectively. These gates are generally designated by latitude-longitude coordinates and an altitude which aircraft must fly through. These gates into and out of N90 TRACON are often subject to closure due to thunderstorm activity. When departure gates are closed, the impacts can be immense: Given the presence of three major airports within N90 TRACON (New York-LaGuardia, New York-JFK, and Newark), closed departure gates result in airport surface congestion, long departure delays, and if severe enough, arrival delays due to lack of airport jet-bridges and taxiway space for incoming arrivals \cite{mukherjee2011flight}. Such delays can have consequences for the entire domestic and international air transportation system.

Given the potential impact of airspace closures, it is imperative to ascertain when weather conditions improve sufficiently to allow for reopening to happen. Given that these arrival and departure gates are tens of thousands of feet in the sky, it is difficult to directly monitor conditions at and around them. However, if weather conditions permit, \emph{exploratory} flights could be directed towards these airspace structures to see if conditions have indeed improved enough to re-open them to all other air traffic. Such flights are known as \emph{pathfinders}, or \emph{pathfinder flights} \cite{FAA2024VP}. Given their outsized importance to air traffic control and flow management, it is important to understand how the decision-making process behind pathfinders can be modeled, then optimized. Additionally, understanding this human-centric behavior can also serve as a baseline for adding automation and autonomy to improve these operations.

\subsection{Previous Works and Research Gap}

The operational value of pathfinders is well established, evidenced by, e.g., dedicated pathfinder working groups within the US Federal Aviation Administration (FAA) \cite{FAA2024Pathfinder}. Additioanlly, terminal airspace operations have been the subject of many previous studies, ranging from flight procedural design \cite{fowler2012getting}, integrating surface-airspace operations \cite{badrinath2019integrated}, and exploring more optimal airspace configurations and geometries \cite{granberg2019framework}. Furthermore, weather-impacted routing within the terminal airspace has also been studied \cite{pfeil2012identification}. Additionally, pathfinders were studied explicitly in \cite{Weber2007}---we aim to provide a systematic approach to modeling and optimizing the pathfinder process. 


\section{Contributions of Work}
\label{sec:contributions}
Our contributions are threefold. First, we develop a Markov chain model that captures the entire pathfinding process under convective weather, enabling steady-state analysis of long-term system dynamics. Second, we propose a decision-making model for the pathfinder selection phase, with stylized representations of flight-level acceptance behavior and controller-level strategies for sequential candidate selection. This model is used in a worst-case analysis to explore how selfless behavior and shared uncertainty affect system resilience. Third, we analyze US FAA coordination logs to illustrate how pathfinder operations are carried out in practice and motivate future data-driven validation of our framework.

\section{Pathfinding Modeling with Markov Chains}
\label{sec:modeling_markov_chain}
We develop a Markov chain model representing the pathfinding process, capturing key operational states and transition probabilities. Steady-state analysis is then performed to examine its long-term operational characteristics.

\subsection{Markov Chain for Pathfinding}
\label{sec:markov_chain}
\begin{figure}[htbp]
  \centering
  \includegraphics[width=\linewidth]{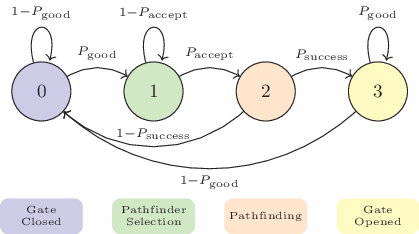}
  \caption{Markov chain representation of the aircraft pathfinding process. The model consists of four states: \textit{Gate Closed} (state 0), \textit{Pathfinder Selection} (state 1), \textit{Pathfinding} (state 2), and \textit{Gate Opened} (state 3). Transition probabilities are governed by \(P_{\mathrm{good}}\), \(P_{\mathrm{accept}}\), and \(P_{\mathrm{success}}\), capturing weather observations, acceptance of pathfinder offers, and pathfinding success, respectively.}
  \label{fig:markov_chain}
\end{figure}
A Markov chain assumes a system's future state depends only on its current state and not on the full history of past states. This memoryless property makes it useful for modeling decision-making under uncertainty. It has been widely used in fields like reliability analysis to model the dynamic behavior of systems over time \cite{markov_chain}. Following this approach, we model the pathfinding process under convective weather using a discrete-time Markov chain, capturing probabilistic transitions between key operational states. Our Markov chain model is illustrated in \Cref{fig:markov_chain}.

We define the Markov chain model with three key probabilistic parameters: \(P_\mathrm{good}\), \(P_\mathrm{accept}\), and \(P_\mathrm{success}\). The probability \(P_\mathrm{good}\) reflects the chance of observing favorable weather, determining whether the system proceeds with pathfinding or remains closed. \(P_\mathrm{accept}\) represents the likelihood that an aircraft accepts the system's request to serve as a pathfinder. \(P_\mathrm{success}\) indicates the probability that the system, based on the pathfinding result, decides to open the gate for subsequent flights.

The system transitions through four states. State~0 denotes the gate being closed due to convective weather, with a transition to State~1 (\textit{Pathfinder Selection}) occurring if weather observations improve (\(P_{\mathrm{good}}\)). In State~1, a candidate flight is evaluated; if the aircraft accepts the mission (\(P_{\mathrm{accept}}\)), the system moves to State~2 (\textit{Pathfinding}), otherwise it remains in State~1 and continues the selection process. A successful pathfinding attempt (\(P_{\mathrm{success}}\)) leads to State~3, where the gate is opened for normal departures, while failure causes the system to revert to State~0. In State~3, the gate remains open if favorable weather persists; if not, the system returns to the closed state.

\subsection{Steady-State Behavior and Long-Term Implications}
\label{sec:steady_state}
We analyze the long-run behavior of the Markov chain through steady-state analysis, which characterizes the limiting distribution over the four operational states defined earlier. The system evolution is governed by the transition matrix \( P \).
\begin{equation}
P = 
\begin{pmatrix}
1 - P_{\mathrm{good}} & P_{\mathrm{good}} & 0 & 0 \\
0 & 1 - P_{\mathrm{accept}} & P_{\mathrm{accept}} & 0 \\
1 - P_{\mathrm{success}} & 0 & 0 & P_{\mathrm{success}} \\
1 - P_{\mathrm{good}} & 0 & 0 & P_{\mathrm{good}}
\end{pmatrix}.
\end{equation}

The steady-state distribution~\( \pi = (\pi_0, \pi_1, \pi_2, \pi_3) \)~can be obtained by solving the balance equation and the normalization condition \cite{MC_solve}, while ensuring that all parameters are valid probabilities \((P_{\mathrm{good}}, P_{\mathrm{accept}}, P_{\mathrm{success}}) \in [0, 1]\).

\begin{equation}
\pi P = \pi, \quad \sum_{i=0}^{3} \pi_i = 1.
\end{equation}

\begin{figure*}[htbp]
  \centering
  \includegraphics[width=\linewidth]{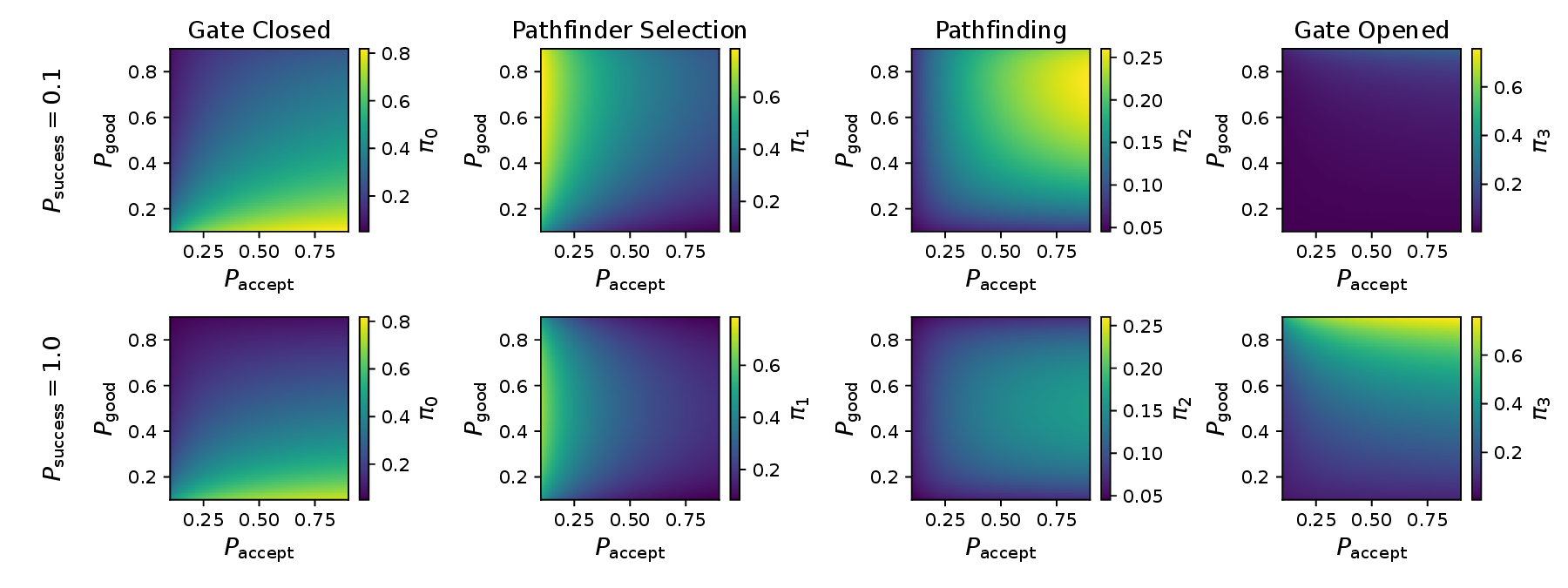}
  \caption{Steady-state distribution (\(\pi_0\) to \(\pi_3\)) of a four-state Markov chain under varying values of \(P_{\mathrm{good}}\) and \(P_{\mathrm{accept}}\), for two different values of \(P_{\mathrm{success}}\): 0.1 (top row) and 1.0 (bottom row). Each column corresponds to one of the four system states: \textit{Gate Closed}, \textit{Pathfinder Selection}, \textit{Pathfinding}, and \textit{Gate Opened}. Color intensity represents the steady-state probability \(\pi_i\) for each state. The same color scale is used within each column for fair comparison between the two \(P_{\mathrm{success}}\) settings.}
  \label{fig:steady_state}
\end{figure*}
We compute this distribution numerically in our analysis. The sensitivity analysis results are shown in \Cref{fig:steady_state}, illustrating how long-run occupancy varies across the four system states under different values of \(P_{\mathrm{good}}\), \(P_{\mathrm{accept}}\), and \(P_{\mathrm{success}}\). The figure compares scenarios with low and high pathfinding success, highlighting how input parameters affect steady-state probabilities.

As \(P_{\mathrm{good}}\) increases, the system tends to remain in the \textit{Gate Opened} state (\(\pi_3\)), while the \textit{Gate Closed} state (\(\pi_0\)) becomes less frequent. Increasing \(P_{\mathrm{accept}}\) leads to more transitions into pathfinding, reflected in lower \(\pi_1\) and \(\pi_2\). A key relationship, \(\pi_2 = P_{\mathrm{accept}} \pi_1\), ensures that pathfinding (State~2) always has lower steady-state probability than \textit{Pathfinder Selection} (State~1), explaining why the system spends relatively little time in State~2—--especially when \(P_{\mathrm{success}} = 1\), where it mostly alternates between \textit{Gate Closed} and \textit{Gate Opened} states.

When pathfinding success is low (\(P_{\mathrm{success}} = 0.1\)), the system becomes dominated by \textit{Gate 
Closed} and selection states, while the \textit{Gate Opened} state (\(\pi_3\)) sharply decreases. The rise in \(\pi_2\) reflects its dependence on \(\pi_1\), and maintaining throughput requires high values of both \(P_{\mathrm{good}}\) and \(P_{\mathrm{accept}}\).

This steady-state analysis highlights that the system's ability to sustain nominal operations depends not only on environmental conditions but also on decision-making parameters, particularly \( P_{\mathrm{accept}}\), which determines how readily the system transitions into pathfinding. As \( P_{\mathrm{accept}}\) increases, the system is more responsive to opening gates, reducing time spent in intermediate states. This observation directly motivates the pathfinder selection modeling in the next section, where we examine how \( P_{\mathrm{accept}}\) emerges from agent- and controller-level decisions. 

\section{Stylized Modeling of Pathfinder Selection}
\label{sec:pathfinder_selection}
This section focuses on stylized modeling within the \textit{Pathfinder Selection} state. We present two complementary perspectives: a flight-centric model that captures agent-level\footnote{We use the term \emph{agent} to refer generally to an atomic decision-making entity; for pathfinders, this is a flight. We chose this general term to reflect the fact that our framework can be applicable for a broader set of decision-making problems.} decision logic, and a controller-centric model that reflects operational priorities in pathfinder assignment.

We model the air traffic controller as a centralized decision-maker that follows a predetermined priority order over candidate flights. Offers are extended sequentially according to this order: if a flight rejects the offer, the controller proceeds to the next flight. If the offer is accepted, pathfinding is initiated. Accordingly, our modeling centers on how flights decide to accept or reject offers, and how the controller determines the offer sequence.

\subsection{Flight-Centric Stylized Representation}
\label{sec:flight_model}
To model pathfinder acceptance behavior at the individual level, we introduce a stylized, flight-centric representation based on agent utility. We consider a set of decision-making agents \( N = \{1, 2, \dots, n\} \), where each agent \( i \in N \) chooses whether or not to participate in pathfinding. Agents are modeled as independent decision-makers, with no coordination or influence between them. Each agent makes a binary decision, defined as
\begin{equation}
x_i =
\begin{cases}
1, & \text{if agent } i \text{ accepts pathfinding offer,} \\
0, & \text{otherwise.}
\end{cases}
\end{equation}

Each agent evaluates its decision to accept or reject the pathfinder offer based on three components: participation cost, failure cost, and reward. The participation cost \( c_i(x_i) \) reflects the operational or perceived burden of accepting the role (i.e., \enquote{how much will pathfinding cost me?}), with \( c_i(0) = 0 \). The failure cost \( d_i(x_i) \) represents the expected loss if the mission fails, and also satisfies \( d_i(0) = 0 \). Here, failure refers to the case where the agent accepts the pathfinder role and departs, but is unable to open the gate because the weather remains unfavorable. Such failed pathfinding may result in fuel penalties, deviations, and increased passenger discomfort due to experienced turbulence. The reward \( T_i(x_i) \) is granted when the agent accepts the offer and departs, regardless of whether or not the pathfinding succeeds. This reflects the agent’s incentive to depart immediately and avoid additional delay, which is treated as compensation.

The agent’s decision determines the probability of success, given by
\begin{equation}
P_{\mathrm{success},\,i}(x_i) =
\begin{cases}
P_{\mathrm{success},\,i}, & \text{if } x_i = 1, \\
0, & \text{if } x_i = 0.
\end{cases}
\end{equation}
The resulting utility for agent \( i \) is expressed as
\begin{equation}
U_i(x_i) =
\begin{cases}
T_i - c_i - (1 - P_{\mathrm{success},\,i})d_i, & \text{if } x_i = 1, \\
0, & \text{if } x_i = 0.
\end{cases}\label{eq:utility}
\end{equation}
This utility function in \eqref{eq:utility} represents the agent’s net perceived benefit: if the agent accepts the offer (\(x_i = 1\)), it receives a reward \(T_i\), pays a participation cost \(c_i\), and faces an expected failure cost weighted by the mission's success probability. If the agent declines (\(x_i = 0\)), the utility is zero.

\begin{figure}[htbp]
    \centering
    \includegraphics[width=0.9\linewidth]{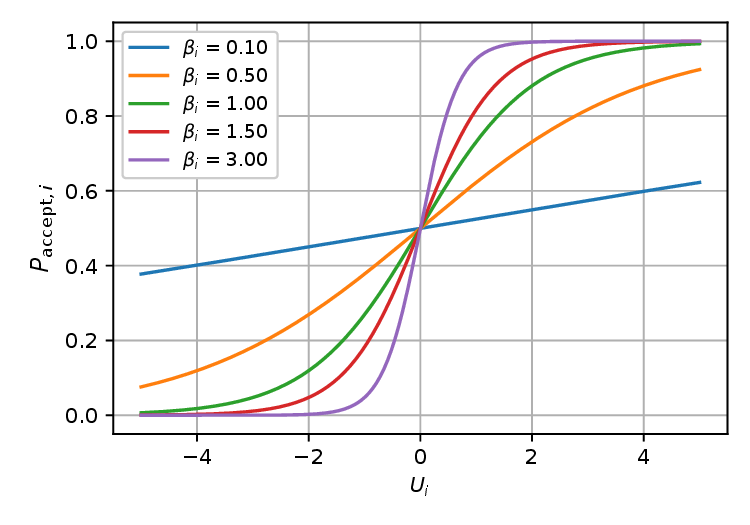}
    \caption{\(P_{\mathrm{accept},\,i}\) as a function of utility \( U_i \) for various values of the sensitivity parameter \( \beta_i \). Higher \( \beta_i \) values make the agent's decision more deterministic, resulting in a sharper transition from rejection to acceptance as utility increases, while lower \( \beta_i \) values cause decisions to be more randomized across a wider range of utilities.}
    \label{fig:flight_utility}
\end{figure}

We model the probability of acceptance using a logistic function, a standard formulation in discrete choice theory that captures bounded rationality through stochastic utility maximization \cite{logit_model}:
\begin{equation}
P_{\mathrm{accept},\,i}(x_i) = \frac{1}{1 + e^{-\beta_i U_i(x_i)}},
\end{equation}
where \( \beta_i > 0 \) controls sensitivity to utility differences: higher values lead to more deterministic choices, while lower values introduce randomness. \Cref{fig:flight_utility} illustrates this behavior.

\subsection{Controller-Centric Stylized Representation}
\label{sec:controller_model}
From the air traffic controller’s perspective, pathfinder selection is a strategic decision to minimize overall system delay. The controller selects a candidate flight based on its expected contribution to reducing system-wide ground delay and the probability of accepting the pathfinding role. The controller balances the trade-off between potential delay reduction and the risk of \emph{rejection}, i.e., a flight refusing to be a pathfinder. 

Accordingly, the controller’s expected payoff is defined as the product of the acceptance probability and estimated delay reduction,
\begin{equation}
\text{Payoff}_{\text{controller}} = P_{\text{accept},\,i} \Delta D_{\text{system},\,i},
\end{equation}
where \( P_{\text{accept}} \) is the probability that the selected flight accepts the offer, and \( \Delta D_{\text{system},\,i} \) denotes the estimated delay reduction if the flight \(i\) successfully completes the pathfinding task.
For agent \( i \), the delay reduction is expressed as,
\begin{equation}
\Delta D_{\text{system},\,i} = \epsilon_i \Delta D_{\text{ideal}},
\end{equation}
where \( \epsilon_i \in [0, 1] \) is a scaling factor that reflects how effective flight \( i \) is in alleviating congestion, and \( \Delta D_{\text{ideal}} \) is the maximum delay reduction achievable by an ideal pathfinder. In practice, system-level impact may depend on contextual factors such as queue position or departure timing. Such considerations motivate using a flight-specific effectiveness factor \( \epsilon_i \).

\section{Worst Case Analysis}
\label{sec:worst_case}
In this section, we investigate a sequential decision process in which agents are individually offered the opportunity to serve as the pathfinder. We focus on the worst-case scenario, where all agents reject the offer, resulting in a complete failure to initiate pathfinding. Each agent probabilistically accepts or rejects the offer based on their perceived utility. Our goal is to understand how the probability of this event changes under different system conditions. We analyze a baseline model and two extensions incorporating selfless behavior and environmental uncertainty. For tractability, we adopt simplifying assumptions explicitly stated in each subsection.


\subsection{Case 1: Under Independent Assumptions}
\label{sec:independent_assumption}
\begin{figure}[htbp]
    \centering
    \includegraphics[width=0.8\linewidth]{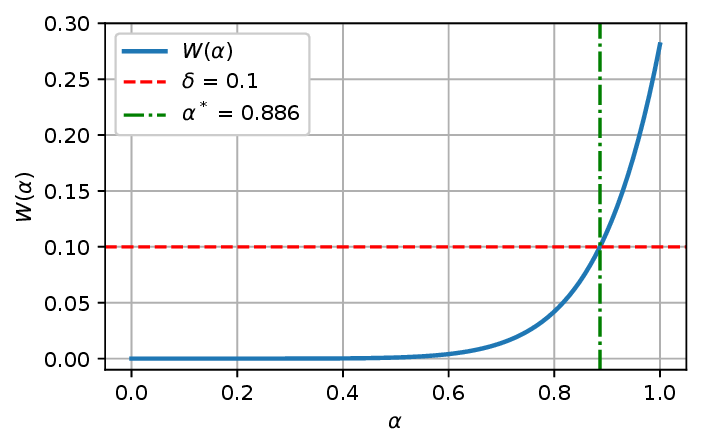}
    \caption{Worst case probability $W(\alpha)$ as a function of the rejective agent ratio $\alpha$. The red dashed horizontal line indicates the system failure threshold $\delta = 0.1$, and the green dash-dotted vertical line shows the critical point $\alpha^*$ where $W(\alpha^*) = \delta$, under $n = 10$, $U^- = -2$, $U^+ = 2$, and $\beta = 1$.}
    \label{fig:vanila_worst}
\end{figure}
We begin with a baseline setting where each agent independently decides to accept or reject the pathfinder offer based on its own utility. Decisions are made without observing others and follow the logistic model in \Cref{sec:flight_model}. The rejection probability is given by
\begin{equation}
P_{\text{reject},\,i}(x_i) = 1 - P_{\text{accept},\,i}(x_i) = \frac{1}{1 + e^{\beta_i U_i(x_i)}}.
\end{equation}
We categorize agents into two groups: rejective and receptive. Each agent is independently assigned to the rejective group with probability \( \alpha \), and to the receptive group with probability \( 1 - \alpha \). We associate these groups with representative utility values \( U^- < 0 \) and \( U^+ > 0 \), respectively, to simplify analysis at the group level. Their rejection probabilities are
\begin{equation}
P_{\text{rejective}} = \frac{1}{1 + e^{\beta U^-}}, \quad
P_{\text{receptive}} = \frac{1}{1 + e^{\beta U^+}},
\end{equation}
with \( P_{\text{rejective}} > 0.5 \) and \( P_{\text{receptive}} < 0.5 \). We assume a common sensitivity parameter \( \beta \) across all agents for simplicity. This setup yields a mixture of agents with differing likelihoods of rejecting the offer.

The worst case scenario occurs when all \( n \) agents reject the offer. Let \( W(\alpha) \) denote the probability of this event, where \( \alpha \in [0, 1] \) represents the fraction of rejective agents in the system. We have that:
\begin{equation}
W(\alpha) = \sum_{k=0}^{n} \binom{n}{k} \alpha^k (1 - \alpha)^{n - k}
(P_{\text{rejective}})^k (P_{\text{receptive}})^{n - k}.
\end{equation}
Using the binomial theorem, this simplifies to
\begin{equation}
W(\alpha) = \left(\alpha P_{\text{rejective}} + (1 - \alpha) P_{\text{receptive}}\right)^n.
\end{equation}
Since \( P_{\text{rejective}} > P_{\text{receptive}} \), \( W(\alpha) \) increases with \( \alpha \), meaning the system becomes more vulnerable as the proportion of rejective agents increases. \( W(\alpha) \) becomes more sensitive to changes in \(\alpha\) when the utilities \( U^- \) and \( U^+ \) are more polarized and the sensitivity \( \beta \) increases, while increasing \( n \) lowers the overall failure probability.

\begin{figure*}[htbp]
    \centering
    \includegraphics[width=\linewidth]{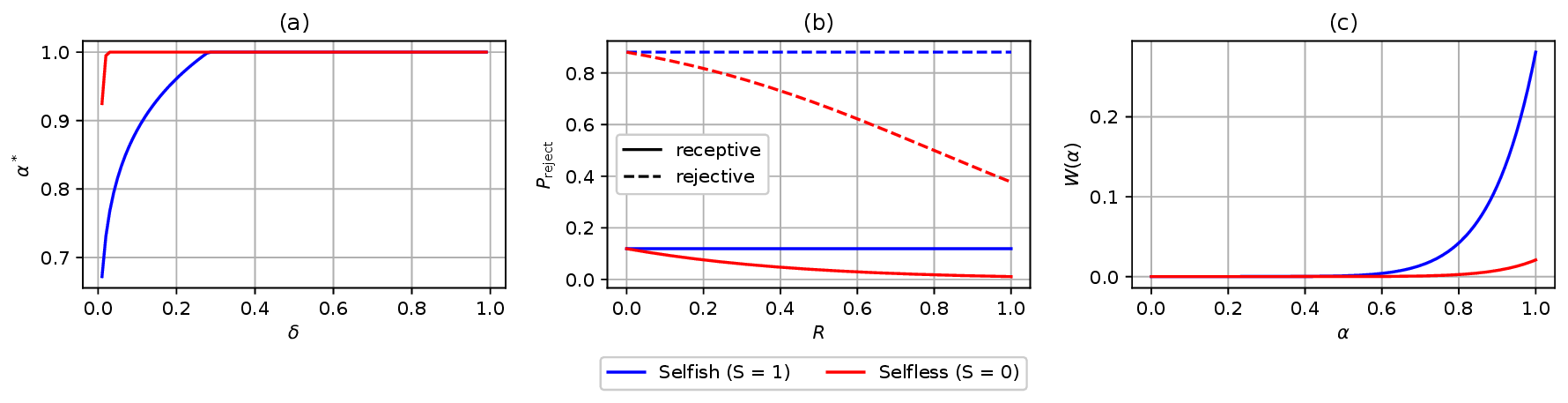}
    \caption{Visualization of worst case analysis across different degrees of selfishness ($S=0$: selfless, $S=1$: selfish) under $n=10$, $U^+ = 2$, $U^- = -2$, $\beta=1$ $\gamma = 2.5$, and $R = 0.5$.
        (a) Tipping point $\alpha^*$ as a function of the system failure threshold probability $\delta$.
        (b) Agent rejection probability $P_{\mathrm{reject}}$ as a function of the observed collective rejection ratio $R$. Solid and dashed lines indicate receptive and rejective groups, respectively.
        (c) Worst case probability $W(\alpha)$ over different $\alpha$ values.
}
    \label{fig:social_worst}
\end{figure*}

We define the critical threshold \( \alpha^* \) such that \( W(\alpha^*) = \delta \) where \( \delta \in (0, 1) \) represents the maximum acceptable probability of complete rejection. Solving the equation yields
\begin{equation}
\alpha^* = \frac{\delta^{1/n} - P_{\text{receptive}}}{P_{\text{rejective}} - P_{\text{receptive}}}.
\end{equation}
This threshold separates stable and fragile conditions. If \( \alpha \geq \alpha^* \) the risk of complete rejection becomes significant, corresponding to system failure. The behavior of \( W(\alpha) \) and the critical threshold \( \alpha^* \) are illustrated in \Cref{fig:vanila_worst} for a specific parameter setting. This analysis offers a clear benchmark for how vulnerable the system becomes when agents act independently. By identifying the tipping point at which widespread rejection causes system failure, operators can proactively adjust selection policies to stay below that risk threshold.

\subsection{Case 2: Considering Agent's Selfless Behavior}
\label{sec:selfless_behavior}
We extend the baseline model to incorporate selfless behavior in pathfinder decision-making. In practice, not all flights evaluate the pathfinder offer solely based on self-interest. Some may consider the broader operational benefit their acceptance could bring, such as enabling departures for other delayed flights. 

To reflect this, we introduce a selfishness parameter \( S_i \) for each agent \( i \), where \( S_i \in [0, 1] \). \( S_i = 1 \) denotes a fully selfish agent who only considers individual utility, while \( S_i = 0 \) corresponds to a fully selfless agent who fully accounts for the system-level impact in their decision. 

The modified utility for agent \( i \) combines their individual utility with a system-level adjustment based on the expected impact of collective rejection.
\begin{equation}
U_i^{\text{sys}} = U_i(x_i) + (1 - S_i) \gamma_i R,
\end{equation}
where \( \gamma_i > 0 \) is a sensitivity parameter, and \( R \) denotes the estimated overall rejection rate in the system. This term reflects the agent's awareness of potential system failure due to widespread rejection, thereby incorporating a notion of social responsibility into the decision.

Based on \( U_i^{\text{sys}} \), the rejection probability becomes 
\begin{equation} 
    P_{\text{reject}}(x_i) = \frac{1}{1 + e^{\beta_i U_i^{\text{sys}}}}. 
\end{equation}

\noindent
The worst case probability under selfless behavior consideration is
\begin{equation}
W_{sys}(\alpha) = \left( \alpha P_{\text{rejective}}^{sys} + (1 - \alpha) P_{\text{receptive}}^{sys} \right)^n,
\end{equation}
where
\begin{align}
P_{\text{rejective}}^{sys} &= \frac{1}{1 + e^{\beta (U^- + (1 - S)\gamma R)}}, \\
P_{\text{receptive}}^{sys} &= \frac{1}{1 + e^{\beta (U^+ + (1 - S)\gamma R)}}.
\end{align}
For tractability, we assume that all agents share the same level of selfishness \( S \) and environmental influence parameter \( \gamma \), which simplifies further analysis. Increased selflessness (lower \( S \)) or stronger concern for system failure (higher \( \gamma \)) reduces rejection probability and improves system resilience. 
\Cref{fig:social_worst} illustrates how selfless behavior improves system robustness. Subfigure~(a) shows that the critical threshold \( \alpha^* \)—defined as the point where the system failure probability \( W(\alpha) \) crosses the threshold \( \delta \)—varies with agent selfishness.  When agents are selfless (\( S=0 \)), the system can tolerate a higher fraction of rejective agents before failure occurs, resulting in a larger \( \alpha^* \). Notably, even for small values of \( \delta \), \( \alpha^* \) approaches one when agents are selfless, indicating high system tolerance. In contrast, under selfishness, \( \alpha^* \) increases much more gradually with \(\delta\), and approaches one only when \(\delta\) becomes sufficiently large. 
In Subfigure~(b), rejection probabilities vary with the estimated rejection rate \( R \) only for selfless agents. As \( R \) increases, selfless agents--—especially those in the rejective group--—become less likely to reject. Subfigure~(c) demonstrates that the overall worst case rejection probability \( W(\alpha) \) is significantly lower under selfless behavior, confirming its positive effect on system resilience. This result helps operators understand how individual rejection behavior scales into system-wide risk, enabling more informed decisions about how many cooperative agents are needed to maintain operational stability.

\subsection{Case 3: Considering Uncertainty in Environments}
\label{sec:environment_uncertainty}
Agent decisions are often influenced by external uncertainty, such as incomplete information or random disturbances. We capture this by adding a stochastic noise term \( \xi \) to the utility function, which we model as a zero-mean random variable with noise level \( \theta \), shared across all agents. This structure reflects the assumption that all agents experience the common realization of external uncertainty. Two different noise models are considered: Gaussian and Rademacher, which allow us to examine the impact of both continuous and discrete randomness on system behavior.

The modified utility incorporates environmental uncertainty as
\begin{equation}
U_i^{\text{env}} = U_i(x_i) + \xi, 
\end{equation}
where \( \xi \) is a random variable with zero mean and scale parameter \( \theta \in \{\sigma, \kappa\} \). For example, \( \theta = \sigma \) if \( \xi \sim \mathcal{N}(0, \sigma^2) \), and 
\( \theta = \kappa \) if \( \xi \sim \text{Rademacher}(\pm \kappa) \). 
A single notation \( \theta \) allows unified treatment across both distributions.

The worst case probability is now a function of both the rejective agent ratio \( \alpha \) and the noise scale \( \theta \); we denote this as \( W(\alpha, \theta). \) The tipping point \( \alpha^*(\theta) \) is implicitly characterized as the value of \( \alpha \) such that \(  W(\alpha^*(\theta),\theta)=\delta \), for a fixed threshold \( \delta \in (0, 1) \).

To apply the implicit function theorem and analyze how \( \alpha^*(\theta) \) changes with respect to \( \theta \), we introduce an auxiliary function
\begin{equation}
G(\alpha(\theta), \theta) := W(\alpha(\theta), \theta) - \delta.
\end{equation}
The condition for the tipping point then becomes \( G(\alpha^*(\theta), \theta)=0 \). This reformulation allows us to differentiate \( \alpha^*(\theta) \) with respect to \( \theta \), treating it as an implicitly defined function. The derivative is given by
\begin{equation}
\frac{\mathrm{d}\alpha^*}{\mathrm{d}\theta} =  - \frac{\partial G / \partial \theta}{\partial G / \partial \alpha}= - \frac{\partial W / \partial \theta}{\partial W / \partial \alpha}.
\end{equation}
This follows from the implicit function theorem applied to \( G(\alpha^*(\theta), \theta) = 0 \). 

Since \( W(\alpha, \theta) \) increases as \( \alpha \) increases—--as a higher rejective agent ratio makes complete rejection more likely—--we have \( \partial W / \partial \alpha > 0 \). Therefore, the sign of \( \mathrm{d}\alpha^* / \mathrm{d}\theta \) is determined by the partial derivative \( \partial W / \partial \theta \), which captures how environmental uncertainty affects the worst case probability. The derivative \( \mathrm{d}\alpha^* / \mathrm{d}\theta \) reflects how the tipping point for failure shifts as the environment becomes more or less uncertain. If the partial derivative \( \partial W / \partial \theta \) is positive, then increasing uncertainty raises the failure probability, causing \( \alpha^* \) to decrease since fewer rejective agents are needed to trigger failure. In this case, \( \mathrm{d}\alpha^* / \mathrm{d}\theta \) is negative, and the system becomes more fragile. Conversely, if \( \partial W / \partial \theta \) is negative, then uncertainty reduces the risk of failure, \( a^* \) increases, and the system becomes more robust to rejection.

\begin{figure}[htbp]
    \centering
    \includegraphics[width=\linewidth]{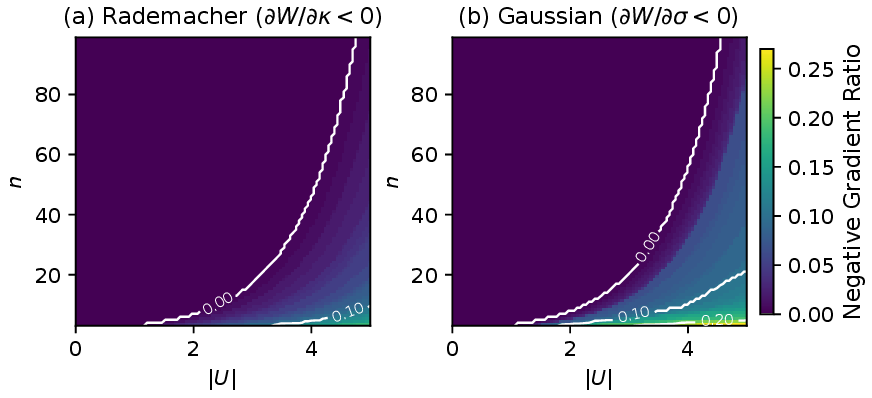}
    \caption{Comparison of gradient behavior under two uncertainty models.  
    (a) Rademacher: fraction of $(\alpha, \kappa)$ pairs with $\partial W / \partial \kappa < 0$.
    (b) Gaussian: fraction of $(\alpha, \sigma)$ pairs with $\partial W / \partial \sigma < 0$.
    Contour lines indicate levels of constant negative gradient ratio.
}
    \label{fig:external_worst}
\end{figure}

\Cref{fig:external_worst} illustrates how the partial derviative \( \partial W / \partial \theta \), which captures how increasing uncertainty affects the worst case probability, varies across the rejective agent ratio \( \alpha \) and the noise scale \( \theta \). Here, \( \theta\) corresponds to the noise parameter: \( \kappa \) for Rademacher noise and \( \sigma \) for Gaussian noise. We evaluate this gradient over a grid of \( (\alpha, \theta) \) values for each instance of number of agents \( n \) and absolute utility \( |U|, \) with \( \alpha \in [0, 1] \) and \( \theta \in [0, 10] \).

The sign of \( \partial W / \partial \theta \) offers insight into whether randomness helps or harms system stability. A positive gradient \( \partial W / \partial \theta > 0 \) indicates that increasing noise increases the probability of complete rejection, making the system more fragile. A negative gradient \( \partial W / \partial \theta < 0 \), by contrast, implies that randomness reduces the chance of failure, making the system more robust.

\begin{table}[htbp]
    \centering
    \caption{Interpretation of the sign of \( \partial W / \partial \theta \).}
    \label{tab:gradient_sign}
    \begin{tabular}{c|p{5.2cm}}
        \hline
        Gradient Sign & Interpretation \\
        \hline
        \( > 0 \) & Uncertainty increases failure (destabilizing) \\
        \( < 0 \) &  Uncertainty decreases failure (stabilizing) \\
        \hline
    \end{tabular}
\end{table}

Across both models in \Cref{fig:external_worst}, we observe that negative gradients are uncommon for most values of absolute utility \( |U| \) and number of agents \( n \). In fact, even when negative gradients do occur across the entire parameter space of \( (\alpha, \theta) \), the fraction of cases with \( \partial W / \partial \theta < 0 \) is generally very low. This suggests that increasing uncertainty \( \theta \) tends to make the system more fragile in most scenarios.

Nevertheless, there are localized regions where the gradient is negative, particularly when \( |U| \) is large and \( n \) is small. Such conditions may occur when agents face unusually strong incentives or penalties (e.g., urgent departures, high operational costs), and when the number of available pathfinder candidates is limited due to low traffic or constrained airspace. The Gaussian case shows slightly more frequent occurrences of negative gradients compared to the Rademacher case. Upon further inspection of the underlying simulation data, we find that negative gradients mostly appear when \( \alpha \) is large, when many agents are predisposed to reject. In such scenarios, introducing randomness can cause just enough agents to deviate from rejection, thereby preventing the system from failing. These observations suggest that noise can stabilize the system when it is already exhibiting a high rejection ratio. In summary, although uncertainty tends to destabilize the system, there are edge cases where it can play a stabilizing role. Understanding when this happens can inform more resilient risk management in uncertain environments.

\section{Integrating with Empirical Data}
\label{sec:dataset}
To ground the proposed modeling framework in real-world operations, we analyze National Traffic Management Log (NTML) data from the FAA which describe pathfinder operations. This section also outlines how the data set is used to support model calibration, validation, and interpretation from a data-driven perspective.

\subsection{Statistical Summary of Dataset}
\label{sec:statistical_summary}
\begin{figure}[htbp]
    \centering
    \includegraphics[width=0.8\linewidth]{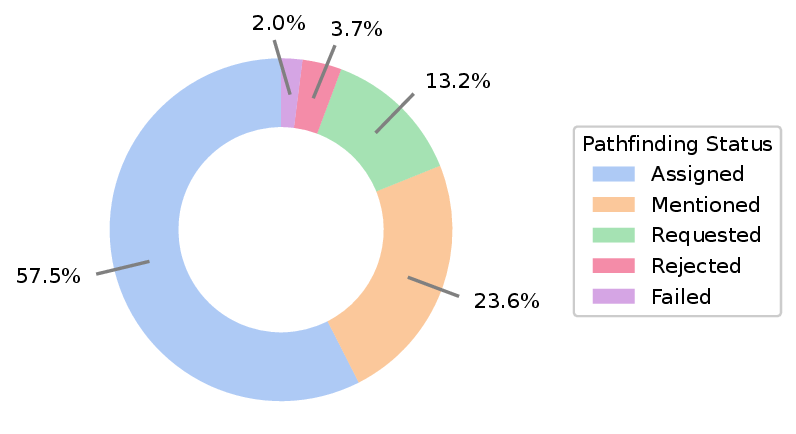}
    \caption{Overall pathfinder comment classification.}
    \label{fig:classification_donut}
\end{figure}

We examine \text{2,178} NTMLs referencing the phrase \enquote{Pathfinder}, collected between \text{December 22, 2022} and \text{December 11, 2024}, across \text{39} different air traffic control facilities. 
We applied a rule-based classification scheme combined with regular expressions to label each comment. The entries were marked as \text{Assigned} if they referenced specific flight numbers (e.g., \texttt{UAL1234}) alongside terms such as ``assigned,'' ``approved,'' ``released,'' or similar expressions. Comments were labeled \text{Requested} when they included intent phrases like ``asking for pathfinder,'' ``can we get one,'' ``requesting,'' or similar language, often without assignment confirmation. The \text{Rejected} label was applied to cases involving coordination denials or pending scenarios, using phrases like ``declined,'' ``no pathfinder,'' ``still waiting,'' or ``not available.'' The \text{Failed} category was reserved for flights that were assigned pathfinding roles but showed poor or ineffective performance, typically indicated by comments stating ``not good,'' ``didn't make it,'' or that the flight deviated from its expected path. Finally, general mentions of the term ``pathfinder'' without flight numbers or action phrases were categorized as \text{Mentioned}. We summarize the distribution of these labels in \Cref{fig:classification_donut}.

\subsection{Steady-State Analysis Using Dataset}
\label{sec:steady_state_results}

Through the NTML data, we conducted a steady-state analysis with Markov chain transition probabilities. Specifically, we calculated fixed \( P_\mathrm{accept} \) and \( P_\mathrm{success} \) using the following formulas:

\begin{equation}
    P_\mathrm{accept} = \frac{N_\mathrm{requested} + N_\mathrm{failed}}{N_\mathrm{requested} + N_\mathrm{failed} + N_\mathrm{rejected}},
\end{equation}
\begin{equation}
    P_\mathrm{success} = \frac{N_\mathrm{requested}}{N_\mathrm{requested} + N_\mathrm{failed}},
\end{equation}
where \( N_\mathrm{requested} \), \( N_\mathrm{failed} \), and \( N_\mathrm{rejected} \) represent the number of comments classified as ``Requested,'' ``Failed,'' and ``Rejected'' in the data set, respectively. Based on the data set, we obtained \( P_\mathrm{accept} = 0.81 \) and \( P_\mathrm{success} = 0.87 \).

\begin{figure}[htbp]
    \centering
    \includegraphics[width=0.93\linewidth]{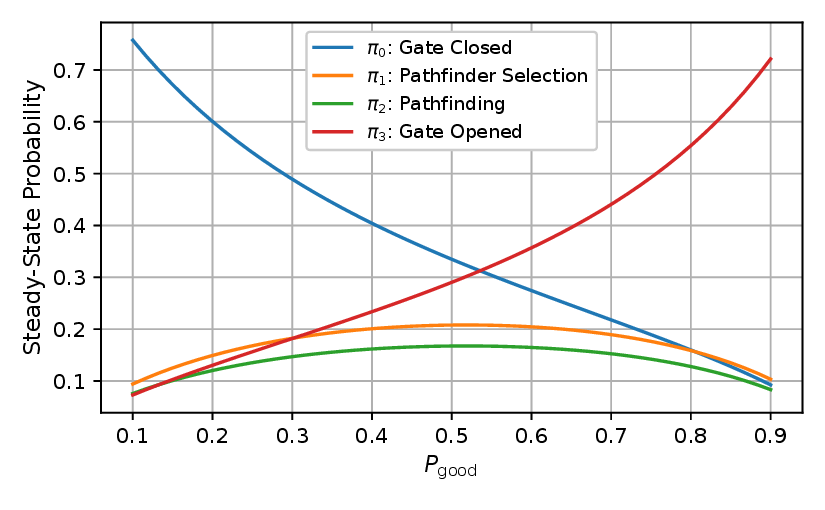}
    \caption{Steady-state probabilities for each system state as a function of gate reliability \( P_\mathrm{good} \), with \( P_\mathrm{accept} = 0.81 \) and \( P_\mathrm{success} = 0.87 \) calculated from NTML data.}
    \label{fig:data_steady_state}
\end{figure}

We computed the steady-state distributions of the Markov chain in \Cref{sec:modeling_markov_chain} using fixed values for \( P_\mathrm{accept} \) and \( P_\mathrm{success} \), while varying the probability of observing favorable weather \( P_\mathrm{good} \). \Cref{fig:data_steady_state} shows how the resulting steady-state probabilities varied with weather reliability, captured by $P_\mathrm{good}$.

As \( P_\mathrm{good} \) increases, the probability of the system remaining in the \textit{Gate Closed} state (\( \pi_0 \)) decreases substantially from 0.75 to 0.09. In contrast, the \textit{Gate Opened} state (\( \pi_3 \)) becomes increasingly dominant, with its steady-state probability rising from approximately 0.07 to 0.72 as \( P_\mathrm{good} \) improves. Meanwhile, the \textit{Pathfinder Selection} (\( \pi_1 \)) and \textit{Pathfinding} (\( \pi_2 \)) states show a modest initial increase followed by a decline, indicating that these are primarily transitional states. These results are consistent with the expected behavior of the Markov chain model and demonstrate how empirical data can be effectively used to parameterize and validate the proposed models.

\subsection{Towards Integrating Data and Stylized Models}
\label{sec:integrating_data}
Building on the steady-state analysis, we aim to further integrate real-world data sources with the stylized decision-making models developed in this study. Our goal is to refine model assumptions and improve alignment with operational behavior observed in practice.

One direction involves combining FAA NTML data with trajectory records from NASA’s Sherlock Data Warehouse \cite{sherlock}. This would enable us to trace how assigned pathfinder flights actually navigated the airspace and evaluate their impact on subsequent traffic flow. By comparing classification labels (e.g., Assigned, Rejected, Failed) with realized flight trajectories, we can more precisely estimate transition probabilities and validate inferred outcomes from our models.

In addition, we aim to calibrate behavioral parameters in the flight-centric stylized decision model, which captures the dynamics of pathfinder request responses. These include reward \( T_i \), participation cost \( c_i \), failure cost \( d_i \), and sensitivity parameter \( \beta_i \) in \Cref{sec:flight_model}. For example, \( \beta_i \) can be estimated for each airline to capture differences in how carriers respond to operational conditions when deciding whether or not to accept a pathfinder role. 

\section{Conclusion}
\label{sec:conclusion}
We present a formal decision-making framework for pathfinder operations in air traffic control under convective weather. Using a Markov chain, we capture the stochastic dynamics of gate accessibility and examine how transition probabilities influence long-term system behavior. Our proposed models describe how individual flights and air traffic controllers make decisions in pathfinder selection, revealing key trade-offs in operational planning. Moreover, our worst case analysis reveals the system's vulnerability to complete rejection of pathfinder requests. We also examine the effects of selfless behavior and environmental randomness on system resilience. Finally, an analysis of NTML data from the FAA confirms the real-world significance of pathfinder operations, highlighting the need for data-driven validation of proposed models.

\subsection{Limitations of Work and Future Directions}

While our models enable analytical tractability, they rely on several simplifying assumptions that limit real-world fidelity. For instance, the worst case analysis assumes independent decision-making across agents, and the selfless behavior and uncertainty models rely on representative shared parameters. The stylized utility formulation also abstracts complex operational incentives into simplified trade-offs. Although model calibration using operational data can help address some of these limitations, capturing more realistic decision-making behavior, such as context-dependent choices, heterogeneous preferences, and adaptive responses, may require more expressive modeling tools. In this direction, insights derived from our current models can inform the development of learning-based methods, such as inverse reinforcement learning or behavioral cloning, to better approximate agent behavior in pathfinder operations.





\section*{Acknowledgments} \label{sec:ack}

\noindent
The authors thank Ron Foley for help with the FAA NTML data set.

\section*{Disclaimer} \label{sec:disc}

\noindent
The contents of this document reflect the views of the authors and do not necessarily reflect the views of the National Aeronautics and Space Administration (NASA), the Federal Aviation Administration (FAA), or the Department of Transportation (DOT). Neither NASA, FAA, nor DOT make any warranty or guarantee, expressed or implied, concerning the content or accuracy of these views.

\bibliographystyle{IEEEtran}
\bibliography{reference}  

\end{document}